\documentclass[pdflatex,sn-aps]{sn-jnl}

\usepackage{graphicx}%
\usepackage{multirow}%
\usepackage{amsmath,amssymb,amsfonts}%
\usepackage{amsthm}%
\usepackage{mathrsfs}%
\usepackage[title]{appendix}%
\usepackage{xcolor}%
\usepackage{textcomp}%
\usepackage{manyfoot}%
\usepackage{booktabs}%
\usepackage{algorithm}%
\usepackage{algorithmicx}%
\usepackage{algpseudocode}%
\usepackage{listings}%
\newcommand {\be}{\begin{equation}}
\newcommand {\ee}{\end{equation}}
\newcommand {\ba}{\begin{eqnarray}}
\newcommand {\ea}{\end{eqnarray}}

\theoremstyle{thmstyleone}%
\theoremstyle{thmstyletwo}%

\theoremstyle{thmstylethree}%

\begin{document}

\title[Article Title]{Non Markovian electron Brownian motion  with radiation reaction force}


\author[1]{\fnm{Juan Francisco} \sur{Garc\'{i}a-Camacho}}\email{jfgarciac@ipn.mx}
\equalcont{These authors contributed equally to this work.}

\author[2]{\fnm{Oliver} \sur{Contreras-Vergara}}\email{ocontrerasv1300@alumno.ipn.mx}
\equalcont{These authors contributed equally to this work.}

\author[2]{\fnm{Norma} \sur{S\'{a}nchez-Salas}}\email{nsanchezs@ipn.mx}
\equalcont{These authors contributed equally to this work.}

\author*[2]{\fnm{Gonzalo} \sur{Ares de Parga}}\email{garesdepargaa@ipn.mx}
\equalcont{These authors contributed equally to this work.}

\author[3]{\fnm{Jos\'{e} In\'{e}s } \sur{Jim\'{e}nez-Aquino}}\email{ines@xanum.uam.mx}
\equalcont{These authors contributed equally to this work.}

\affil[1]{\orgdiv{Departamento de Matem\'aticas}, \orgname{Unidad Profesional Interdisciplinaria de 
Energ\'ia y Movilidad, Instituto Polit\'ecnico Nacional}, \orgaddress{\street{UP Zacatenco}, \city{G. A. Madero}, \postcode{07738}, \state{CDMX}, \country{Mexico}}}

\affil*[2]{\orgdiv{Physics Department}, \orgname{Escuela Superior de F\'{i}sica y Matem\'{a}ticas, Instituto Polit\'ecnico Nacional}, \orgaddress{\street{UP Zacatenco}, \city{G. A. Madero}, \postcode{07738}, \state{CDMX}, \country{Mexico}}}
\affil[3]{\orgdiv{Departamento de F\'isica}, \orgname{Universidad Aut\'onoma Metropolitana-Iztapalapa}, \orgaddress{\street{Apartado Postal 55-534}, \city{Iztapalapa}, \postcode{09340}, \state{CDMX}, \country{Mexico}}}


\abstract{In this work, we study non-Markovian electronic plasma diffusion from a classical point of view, taking into account the effects of the radiation reaction force. The electron Brownian motion is described by a Generalized Langevin Equation (GLE) characterized by an Ornstein-Uhlenbeck-type friction memory kernel. To take into account the effects of the radiation reaction force, an effective memory time which accounts for the thermal interaction of the Brownian particle with its surroundings is proposed. This effective memory time is defined as $\tau_{ef}=\tau-\tau_0>0$, where the memory time $\tau$ accounts for the collision time between electrons in a Brownian motion-like manner, and $\tau_0$ is due to the interaction with the radiation reaction force. Under these conditions, the GLE can be transformed into a stochastic Abraham-Lorentz-like equation, which is analytically solved without violation of causality. The theoretical results will be compared with the numerical simulation.
}




\maketitle

\section{Introduction}\label{sec1}

The Abraham-Lorentz equation \cite{Lorentz1892,Abraham1905} is related to the study of electron dynamics with the radiation reaction force. It was derived from Newton's second law and because of
an additional term proportional to the electron's acceleration rate of change, it represents a classical non-relativistic third-order time derivative equation. This equation leads to paradoxical solutions such as runaway solutions and violation of causality (preaccelerations) \cite{Rohrlich65}. 
The solution of such inconsistencies goes back to the works reported in the context of classical \cite{Dirac1938,DeWitt1960,Landau1975,Rohrlich2000} and quantum 
\cite{Ford1988,Ford1991,Krivitskiui1991,connell2012} electrodynamics. In previous studies \cite{Ford1988,Ford1991,connell2012} electronic plasma diffusion was considered a fluctuation-dissipation phenomenon, described by quantum GLE associated with an electron embedded in a heat bath. 

In a recent paper \cite{Aresdeparga2022}, using classical GLE, it was shown that the effects of the radiation reaction force can be neglected. According to the data reported in \cite{Aresdeparga2022} it is shown that, in a classical non-relativistic regime, the collision time $\tau$ is greater than the electron's characteristic time $\tau _{e}$ and thus the effective memory time $\tau _{ef}=\tau -\tau _{e}\simeq \tau$. Here $\tau$ is of the order of magnitude of the collision time between electrons coming from a Brownian motion-like manner, and $\tau_e=6.26\times 10^{-24}$ $s$, arises owing to the radiation reaction force. Therefore, in this classical description, the effective friction force $m(\tau-\tau_e)\dddot x$, appearing in the stochastic Abraham-Lorentz-like equation (SALE), must be $m(\tau-\tau_e)\dddot x \approx m\tau\dddot x$, therefore, the radiation reaction force $m\tau_e \dddot x$, does not have any effect on the electronic plasma diffusion. 

In the present study, we suggest an alternative proposal by means of the definition of an effective memory time $\tau_{ef}=\tau-\tau_0>0$, 
which allows us to quantify the thermal interaction between a tagged Brownian particle and its surroundings. Here, $\tau_0$ is defined as the memory time resulting from the thermal interaction between the BP and the radiation reaction force. In our proposal, it will be shown that, if $\tau_0=\tau_e$, the effects of the radiation reaction force on the electron Brownian motion become practically imperceptible, as was shown in \cite{Aresdeparga2022}. However, to take into account the effects of the radiation reaction force, we will show that not only the condition $\tau>\tau_0$ must be satisfied, but also some additional condition must be required for other parameters. Our proposal relies upon the
validity of the second Fluctuation-Dissipation Theorem (FDT) established by Kubo in 1966, in the context of the GLE. However, in 1973, the GLE and Kubo's second FDT, were obtained microscopically from the classical statistical mechanics point of view by Zwanzig \cite{Zwanzig1973,Zwanzig2001}. 

On the other hand, it was established in \cite{luczka2005} that in the thermodynamic limit, the memory kernel should tend to a decaying function of time. The
form of this function in the continuous limit depends on the spectral density of the oscillators, defined by $\rho(\omega)$. If $\rho(\omega)\sim \omega$, then the friction memory kernel reduces to Dirac's Delta function, and thus the GLE to the ordinary Markovian Langevin equation. However, if $\rho(\omega)\sim 1/(\omega^2 +\tau^{-2}_{ef})$, in this case it can be shown that the friction memory kernel is proportional to $e^{-t/\tau_{ef}}$, which is precisely an Ornstein-Uhlenbeck-type (OU) process. Due to this fact, the stochastic dynamics associated with the thermal noise appearing in the GLE satisfies the OU-type process.  Once this is done and using Kubo's second FDT, the SALE containing an effective radiation reaction force $m\tau_{ef}\dddot x>0$, can be derived from the GLE. In this case, the solution of the SALE cannot violate causality.

From the SALE, three analytical expressions are explicitly calculated for the variance in position and velocity, depending on the discriminant value $\sqrt{1-4J_{rad}}$, associated with two of the three roots (the third root is zero) of the homogeneous SALE. As it will see later, $J_{rad}$ is defined as $J_{rad}=\tau_{ef}/\tau_r$,  where  $\tau_r=m/\gamma_0$ is the relaxation time and $\gamma_0$ is the friction coefficient. It is clear that in the absence of the radiation reaction force $\tau_0=0$, and thus $J_{rad}$ becomes the usual non-Markovian Brownian motion, that is,  $J_{rad}=J_{bm}=\tau/\tau_r$. Therefore,  the difference $\Delta J\equiv J_{bm}-J_{rad}=\tau_0/\tau_r$, must be the dimensionless parameter that determines when the effects of the radiation reaction force can be taken into account. This can be analyzed in the following way: 

First of all, it will be shown that in the case of the critical root, $J_{rad}=J_{bm}\equiv J_c=1/4$, such that $\tau=\tau_{ef}$, it has that $\tau_0=0$ and $\Delta J=0$. This means that the critical case could represent very well the theoretical description given in Ref. \cite{Aresdeparga2022}, in the following sense: if the time-scale $\tau_0=\tau_e\sim 10^{-24}$ $s$, then $\Delta J\sim \tau_e$, which is practically zero, and therefore, there is no influence of the radiation reaction force on the electron plasma diffusion. It must be evident that for the case of real roots, for which $J_{rad}<1/4$, the effects of the radiation reaction force are not noticeable either. So that, for values of $J_{rad}$ less or equal to the  critical value $J_c=1/4$, the effects of the radiation reaction force are imperceptible. Due to this fact, the only possibility of taking into account such effects must be in the case of complex roots, such that $J_{rad}>J_c=1/4$. More precisely, it will be shown that the influence of the radiation reaction force on the electron plasma diffusion should be above the threshold value ${\mathcal J}=3$. Our work adds to the list of works reported in the literature related to the study of Brownian motion with radiation reaction force within Classical and Quantum Mechanics, and considers gravitational effects \cite{Hsiang2022,Hsiang2019,Bravo2023,Galley2005,Galley2006,Johnson2002}.

The remainder of this paper is organized as follows. In Sec. II, a brief study on Zwanzig's model related to the microscopic derivation of the GLE as well as the derivation of the SALE is
given. Section III focuses on the analytical solution of the SALE to obtain explicit expressions for the Mean Square Displacement (MSD) and Mean Square Velocity (MSV). The analytical results are compared with the numerical simulation of the SALE, to validate the consistency of the theoretical results. The conclusions are presented in Sec. IV.

\section{GLE and the Stochastic Abraham-Lorentz-like equation}\label{sec2}

\subsection{Zwanzig's model }
The system of interest is a tagged Brownian particle of electric charge $Q$ and mass $m$, embedded in a fluid (the bath) constituted of a large number of independent particles acting as harmonic oscillators, each with mass $m_i$, charge $q_i$, and frequency $\omega_i$. In Zwanzig's model, the tagged Brownian particle (BP) is linearly coupled to the bath oscillators through a constant coupling parameter $c_i$; however, we will suppose that $c_i$ also takes into account the thermal interaction between the BP and the radiation reaction force. The momenta for the tagged and bath particles are $p$ and $p_i$, respectively. 
According to Zwanzig's model, the Hamiltonian of the total system becomes 
\begin{eqnarray}
H={p^2\over 2m}+U(x)
+\sum_i \left[{p^2_i\over 2m_i}+{m_i\omega_i^2\over 2}\left(x_i -{c_i\over\omega^2_i} x\right)^2\right] ,           \label{He}  \end{eqnarray}
where $U(x)$ accounts for the potential energy associated with some linear or non-linear force.   According to Hamilton's equations, we get the set of differential equations. 
\begin{eqnarray}
\dot x&=& {p\over m}, \qquad \dot p= -U^{\prime}(x)+ \sum_i m_i c_i \left(x_i - {c_i\over \omega_i^2} x\right)  , \label{dxdp} \\
 \dot x_i &=& {p_i\over m_i}, \qquad \dot p_i=-m_i\omega_i^2 x_i + m_i c_i x  .  \label{dxidpi} 
\end{eqnarray}
Using the Laplace transform to solve the differential equation for $\ddot x_i$, and the subsequent algebra, we arrive to the GLE \cite{Kubo1966,Zwanzig2001,luczka2005} 
\begin{equation}
m \dot v(t)=-U^{\prime}(x) -\int_0^t \gamma(t-t^{\prime})  \, v(t^{\prime}) \, dt^{\prime}  +f(t) ,  \label{gle1} 
\end{equation}
where $\gamma(t-t^{\prime})$ is the generalized friction memory kernel given by
 \begin{equation}
 \gamma(t-t^{\prime})= \sum_i  {m_ic^2_i \over\omega_i^2} \cos[\omega_i (t-t^{\prime})], \label{g}
\end{equation}
and $f(t)$ the thermal noise which reads as 
\begin{eqnarray}
 f(t)=\sum_i \bigg\{  m_i c_i \left(x_i(0) -{c_i \over \omega^2_i} x(0)\right) \cos(\omega_i t) 
 +  {c_i\over \omega_i} p_i(0) \sin(\omega_i t) \bigg\} .   \label{f}
\end{eqnarray}
Following Zwanzig's method \cite{Zwanzig2001}, it can be shown that $\langle f(t)\rangle=0$ and the noise correlation function satisfies  
\be
\langle f(t) f(t^{\prime})\rangle=k_BT\gamma(t-t^{\prime}) , \label{Kfdt} 
\ee
which is precisely Kubo's second FDT. An exponentially decaying function for the friction memory kernel can be obtained in the thermodynamic limit for which the spectral density of the harmonic oscillators is defined by \cite{luczka2005}
 \begin{equation}
 \rho(\omega)=\sum_i {m_ic^2_i\over \omega_i} \delta(\omega-\omega_i) ,
  \label{rho}
\end{equation}
which allows one to write, in the continuous limit, the memory function (\ref{g}) as
 \begin{equation}
 \gamma(t) =\int_0^{\infty} {\rho(\omega)\over\omega} \cos(\omega t) \, d\omega . \label{ga1}
\end{equation}
If the spectral density satisfies $\rho(\omega)= {2\gamma_0 \over \pi\tau_{ef}^2}\, {\omega\over  \omega^2+\tau_{ef}^{-2}}$, with $\tau_{ef}=\tau-\tau_0$, then the memory function becomes  
 \begin{equation}
 \gamma(t)= {2\gamma_0 \over \pi \tau_{ef}^2}   
 \int_0^{\infty}  { \cos(\omega t) \over \omega^2+\tau_{ef}^{-2}} \, d\omega = 
 {\gamma_0 \over \tau_{ef}}  e^{-t/\tau_{ef}} ,  \label{ga2}
\end{equation}
 and therefore 
 \begin{equation}
\gamma(t-t^{\prime})=   {\gamma_0\over\tau_{ef}} e^{-|t-t^{\prime}|/\tau_{ef}} ,   \label{gou}
\end{equation}
whether $t>t^{\prime}$ or $t<t^{\prime}$. According to Eq. (\ref{Kfdt}), the FDT in this case becomes 
\be 
\langle f(t)f(t^{\prime} \rangle ={\gamma_0 k_B T\over\tau_{ef}}\,  e^{-|t-t^{\prime}|/\tau_{ef}} . \label{fdt1}
\ee
This expression suggests that the thermal noise $f(t)$ satisfies the OU-type process such that  
\be
 \dot f(t)=-{1\over \tau_{ef}} f +{\sqrt{\lambda}\over\tau_{ef}}\, \xi(t) ,
  \label{df} \ee
where $\xi(t)$ is Gaussian white noise with zero mean value and correlation function $\langle \xi(t)\xi(t^{\prime}) \rangle= 2\, \delta(t-t^{\prime})$, and $\lambda=\gamma_0 k_B T$. The solution of Eq. (\ref{df}) reads 
\begin{equation}
 f(t)=  f(0) e^{-t/\tau_{ef}}+
 {\sqrt{\lambda}\over \tau_{ef}} \int_0^t e^{-(t-t^{\prime})/\tau_{ef}} \, \xi(t^{\prime}) \, dt^{\prime} . \label{fa}  
\end{equation}
By assuming that $\langle f(0)\xi(t)\rangle =0$, it can be shown that, after some algebra, the noise correlation function satisfies 
\begin{equation}
\langle f(t)f(t')\rangle= \langle f^2(0)\rangle  e^{-(t+t')/\tau_{ef}}
+{\lambda \over \tau_{ef}} [e^{-|t-t'|/\tau_{ef}} - e^{-(t+t')/\tau_{ef}}] . \label{ncf-a}  
\end{equation}
For values of time $t\gg 1/\tau_{ef}$, ~$t'\gg \tau_{ef}$, we can conclude that  
\begin{equation}
 \langle f(t) f(t')\rangle= {\gamma_0 k_BT\over \tau_{ef}} e^{-|t-t'|/\tau_{ef}} .  \label{fdt3}  
\end{equation}

\subsection{Stochastic Abraham-Lorentz-like equation}

The memory function by Eq. (\ref{gou}) allows us to obtain the SALE for an electron gas performing a Brownian motion, taking into account the radiation reaction force. In this case, the GLE (\ref{gle1}) without the potential force $U^{\prime}(x)$ is written as 
\be
 m\dot v=-{\gamma_0\over\tau_{ef}}\int_0^t \gamma(t-t^{\prime})\, v(t^{\prime})\, dt^{\prime} + f(t) , \label{gle2} 
 \ee
So, taking into account Eqs. (\ref{df}) and (\ref{fa}), the GLE (\ref{gle2}) can be transformed into a third-order time-dependent stochastic differential equation.
\be m \tau_{ef} {\dddot x} +m{\ddot x}+\gamma_0 \dot x= \sqrt{\lambda} ~\xi(t) , \label{sale} \ee
which is precisely the SALE associated with electronic plasma diffusion, when this system is influenced by the radiation reaction force quantified by the friction force $m \tau_{ef} {\dddot x}$. We have assumed that $\tau_{ef}>0$ and therefore the solution of the SALE is free of paradoxical solutions.

\section{Explicit solution of SALE for the MSD and MSV}\label{sec3}

In this section, we obtain the analytical solutions of the MSD and MSV by means of the explicit solution of the SALE. First of all, the  homogeneous solution of Eq. (\ref{sale}) has three roots given by   
\be
\lambda_1=0, \qquad \lambda_2=-{1\over 2\tau_{ef}}+{1\over 2\tau_{ef}}\sqrt{1-4J_{rad}}, \qquad \lambda_3=-{1\over 2\tau_{ef}}-{1\over 2\tau_{ef}}\sqrt{1-4J_{rad}} , \label{l23}  \ee
from which, three cases can be analyzed, namely, the real ($J_{rad}<1/4)$, critical ($J_{rad}=1/4$), and complex  ($J_{rad}>1/4$) roots, where $J_{rad}=\tau_{ef}/\tau_r=(\tau-\tau_0)/\tau_r$, and $\tau_r=m/\gamma_0$ is the relaxation time. In the three cases, we explicitly calculate the variance in position and velocity assuming zero initial conditions for the mean values $\langle x(0)\rangle=0$ and $\langle v(0)\rangle=0$.
So, the position variance or Mean Square Displacement (MSD) reads
$\sigma_x^2(t)\equiv\langle x^2(t)\rangle-\langle x(t)\rangle^2=\langle x^2(t)\rangle$, and the velocity variance or Mean Square Velocity (MSV) reads $\sigma_v^2(t)\equiv\langle v^2(t)\rangle-\langle v(t)\rangle^2=\langle v^2(t)\rangle$.

\subsection{Critical root} 

For the reasons given in the introduction, we begin our study with the critical root for which $J_{rad}=J_{bm}\equiv J_c=1/4$ and $\tau_{ef}=\tau=\tau_r/4$. In this case the MSD and MSV become 
\ba 
\langle x^2(t) \rangle&=& 2Dt-\tau D\bigg[
11 - \left({t\over\tau}+4\right)e^{-{t\over 2\tau}} + {1\over 2}
\bigg({t^2\over\tau^2} + 6{t\over\tau} +10\bigg) e^{-{t\over\tau}} \bigg] , \label{cx2}\\
\nonumber\\
\langle v^2(t) \rangle&=&{k_{_B}T\over m}\bigg[1-{1\over 2}\bigg( {t^2\over\tau^2}+2{t\over\tau} +2\bigg)e^{-{t\over\tau}} \bigg]. \label{cv2}
\ea
In the long time limit, such that $t\gg \tau$
\be 
\langle x^2(t)\rangle =2D\left(t - {11\over 8} \tau_r\right), \qquad 
\langle v^2\rangle_{eq}={k_BT\over m}.  \label{cx23} \ee
where $D=\lambda/\gamma_0^2=k_BT/\gamma_0$ is the Einstein diffusion coefficient. Hence, in the critical case, the radiation reaction force has no influence on the electron plasma diffusion, as can be corroborated in the MSD and MSV given by Eq. (\ref{cx2}) and (\ref{cv2}). Moreover, in the long time limit, the MSD is the same as the standard Markovian diffusion process, except by a time delay $t-{11\over 8}\tau_r$.   
  
\begin{figure}
    \centering
    \includegraphics[width=0.55\linewidth]{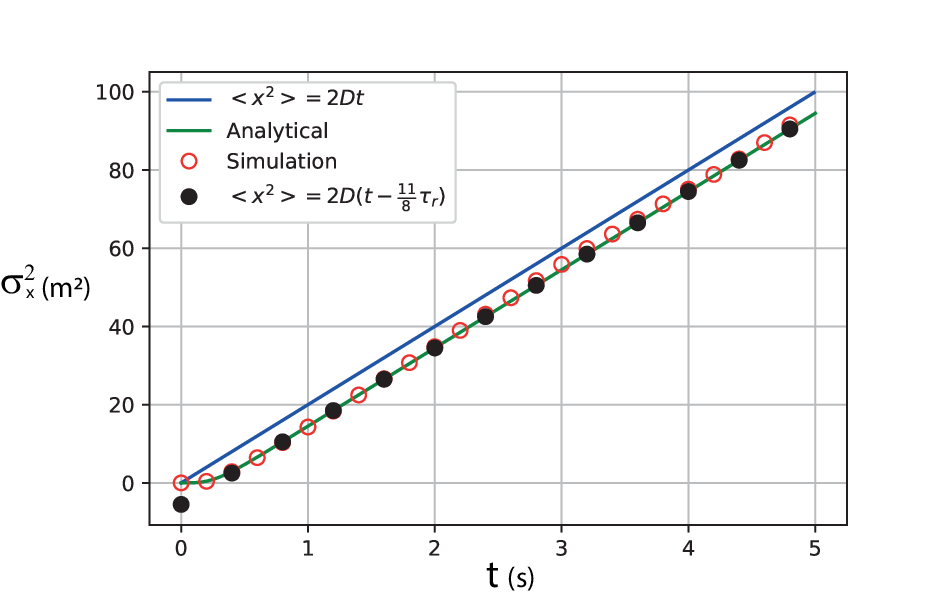}
    \caption{MSD in the case of critical root. Blue line corresponds to Markovian MSD. Green line is the analytical result (\ref{cx2}), and red circles the simulation results of Eq. (\ref{sale}). Black dots  
    are related to long time MSD (\ref{cx23}).
    The used parameters are: $J_c=0.25, ~~\gamma_0=5, ~~\tau=0.05, ~~\tau_r=0.20$.}
    \label{Fig.1}
\end{figure}

\subsection{Real roots} 

\begin{figure}
    \centering
    \includegraphics[width=0.55\linewidth]{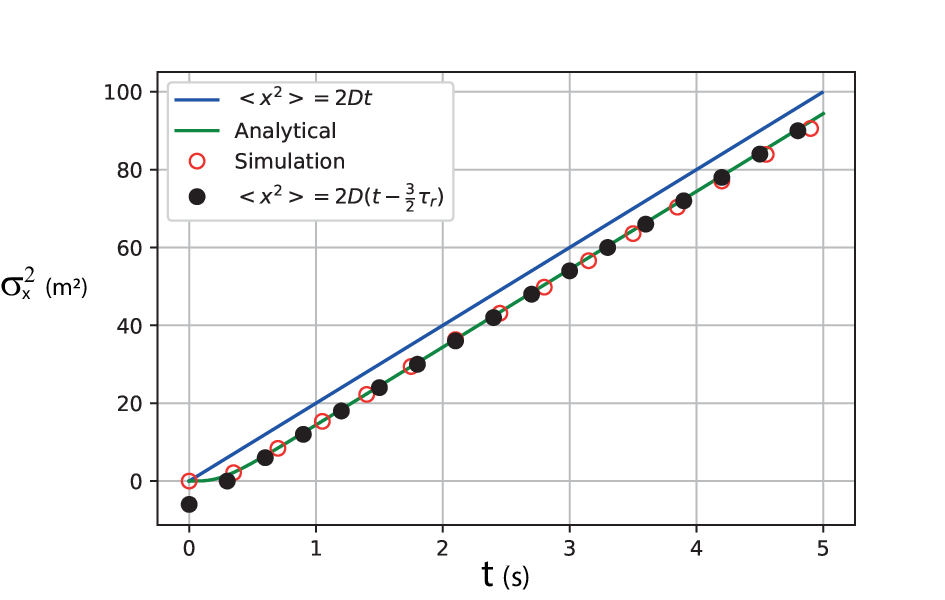}
    \caption{MSD in the case of real roots. Blue line corresponds to Markovian MSD. Green line  is the analytical result (\ref{rx2}), and 
    red circles the simulation result of Eq. (\ref{sale}).
   Black dots are related to long time MSD (\ref{rx23}). The used parameters are:  $J_{rad}=0.20, ~~\gamma_0=5, ~~\tau_{ef}=0.04, ~~\tau_r=0.20$.}
    \label{Fig.2}
\end{figure}

According to the aforementioned initial conditions, the MSD and MSV are explicitly given by
\ba
\langle x^2(t)\rangle&=&2Dt+ J_{rad}^2 {D \over {\tau_{ef}^4}} \bigg\{{e^{2\lambda_2\, t}-1 \over \lambda_2^3 (\lambda_3-\lambda_2)^2} +{e^{2\lambda_3\,t}-1 \over \lambda^3_3 (\lambda_3-\lambda_2)^2} \cr
&+&4\bigg[ {e^{\lambda_3\, t}-1\over \lambda_2\lambda_3^3(\lambda_3-\lambda_2)}-{e^{\lambda_2\, t}-1\over \lambda_3\lambda_2^3(\lambda_3-\lambda_2)}-{e^{(\lambda_3+\lambda_2)\, t}-1\over \lambda_3\lambda_2(\lambda_3-\lambda_2)^2(\lambda_3+\lambda_2)} \bigg] \bigg\},
\label{rx2}  \ea
\ba
\langle v^2(t)\rangle
={J_{rad}\, k_{_B}T\over m{\tau_{ef}^3}(\lambda_3-\lambda_2)^2} \bigg[{e^{2\lambda_2\, t}-1\over \lambda_2}-4{e^{(\lambda_3+\lambda_2)\, t}-1\over \lambda_3+\lambda_2}+{{e^{2\lambda_3\, t}-1}\over{\lambda_3}} \bigg], \label{rv2}  \ea
 It can be shown that, in the long time limit, $t\gg \tau_{ef}$, the MSD and MSV now become   
\be 
\langle x^2(t)\rangle=2D\left(t+{J_{rad}-3 \over 2}\tau_r\right) , \qquad  \langle v^2\rangle_{eq}={k_BT\over m}.
\label{rx22} \ee
 It should be noted that in the absence of the radiation reaction force, the MSD and MSV are, respectively, the same as Eqs. (\ref{rx2}) and (\ref{rv2}), with $\tau_{ef}$ replaced by $\tau$ and  $J_{rad}$ by $J_{bm}$. Also, in the long time limit, the MSD is the same as Eq. (\ref{rx22}), except that $J_{rad}$ must be replaced by $J_{bm}$. However, in both cases 
the value of $J_{rad}=J_{bm}=3$ is prohibited because, in each case $0<J<0.25$, which means that $J_{rad}-3\approx -3$ or $J_{bm}-3\approx -3$ and thus 
\be \langle x^2(t)\rangle =2D\left( t- {3\over 2}\tau_r\right). \label{rx23}
\ee
So that, in the long time limit, the MSD is clearly independent of $J_{rad}$ or $J_{bm}$ parameter. Although the MSD (\ref{rx2}) is a function of $J_{read}$, the influence of this parameter on its evolution does not play a role. This fact  can be corroborated in Fig. \ref{Fig.2} where the evolution of the MSD is plotted for the value of $J_{rad}=0.2$. Moreover, as can be seen in the same plot, the evolution of the MSD rapidly behaves as a standard Markovian diffusion process (straight line), except for a time delay $t-{3\over 2}\tau_r$. The analytical results are validated with the numerical simulation of Eq. (\ref{sale}), as also shown in Fig. \ref{Fig.2}. In summary, in the case of real roots such that $J_{rad}<0.25$, the radiation reaction force does not play a relevant role either.

\subsection{Complex roots} 

\begin{figure}
    \centering
    \includegraphics[scale=0.55]{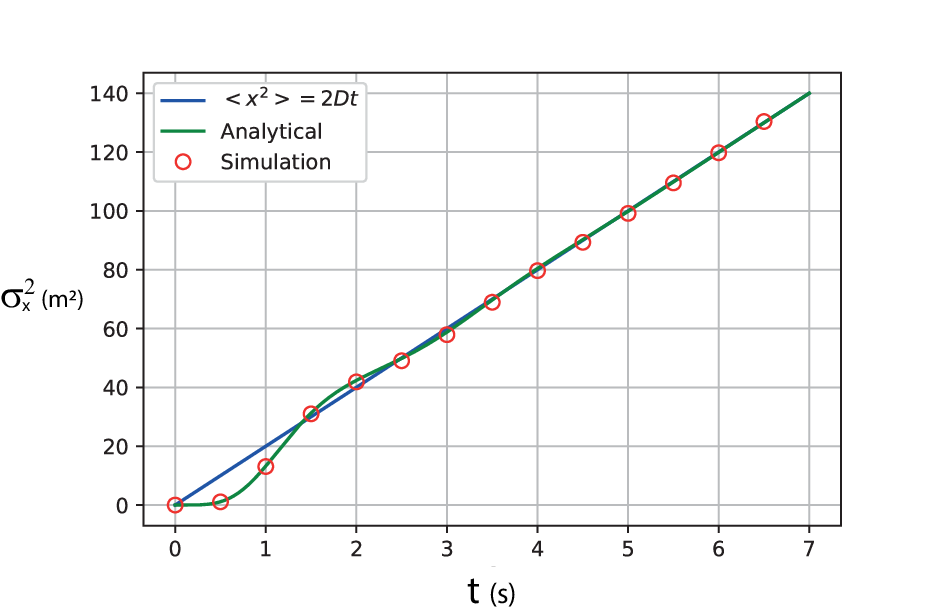}
    \caption{MSD in the case of complex roots. 
    Blue line corresponds to Markovian MSD. 
    Green line is the analytical result (\ref{x2c}) with $J_{rad}=3$, and red circles the numerical simulation of Eq. (\ref{sale}). The used parameters are $\gamma_0=5$, ~~$\tau=0.60$.}
    \label{Fig.3}
\end{figure}

In the two previous sections, we have shown that for $J\le 0.25$, the electronic plasma diffusion is not influenced by the radiation force. In this section we pay attention to the case of complex roots such that $J_{rad}>0.25$, and show under what conditions it is possible to take into account the radiation reaction force on the electronic plasma diffusion. In this case, $\Delta J=J_{bm}-J_{rad}$, should play a relevant role. For this case, the MSD and MSV are given by 
\ba 
\langle x^2(t) \rangle&=&2Dt+
\tau_r D\big[ K_1^2\, {\mathcal B} + K_2^2\, {\mathcal C} - 2\big(K_1\, {\mathcal D} - K_2\,{\mathcal E} + K_1K_2\, {\mathcal F}\big) \big], \label{x2c} \\ 
\nonumber\\
\langle v^2(t)\rangle &=& \tau_r^2\,{k_{_B}T\over m}
\big[\dot K_1^2\, {\mathcal B} -2 \dot K_1 \dot K_2 \, {\mathcal F} +\dot K_2\,{\mathcal C}\big] , \label{v2c}
\ea
where 
\be
{\mathcal B}=-{\tau_{ef}^2}b^2-e^{{t\over\tau_{ef} }}\big[{\tau_{ef}^2}a^2\cos(2b\, t) -{\tau_{ef}^2}ab\sin(2b\,t) -J_{rad} \big],  \label{mB} \ee
\be
{\mathcal C}=-{\tau_{ef}^2}a^2-J_{rad}+e^{{ t\over\tau_{ef}}} \big[{\tau_{ef}^2}a^2\cos(2b\,t)-{\tau_{ef}^2}ab\sin(2b\,t) +J_{rad}\big],  \label{mC} 
\ee
\be
{\mathcal D}=2{\tau_{ef}}b -2e^{t \over 2\tau_{ef}} \big[{\tau_{ef}}a\sin(b\,t) +{\tau_{ef}}b\cos(b\,t) \big],  \label{mD} \ee
\be
{\mathcal E}= 2{\tau_{ef}}a -2e^{t\over 2\tau_{ef}}\big[{\tau_{ef}}a\cos(b\,t) -{\tau_{ef}} b\sin(b\,t) \big],  \label{mE} \ee
\be
{\mathcal F}={{\tau_{ef}}b\over 2}- {1\over 2} e^{t \over \tau_{ef}}\big[{\tau_{ef}}a\sin(2b\,t) +{\tau_{ef}}b\cos(2b\,t)\big].  \label{mF} \ee
\be
K_1=e^{-{t\over 2\tau_{ef}}}\bigg[\frac{a}{b} \cos(b\,t) + \sin(b\,t) \bigg], \label{mK1} \ee

\be
K_2=e^{-{t\over 2\tau_{ef}}}\bigg[\frac{a}{b} \sin(b\,t) -\cos(b\,t) \bigg] . \label{mK2}
\ee
Being $\dot K_1$ and $\dot K_2$ the time derivative of $K_1$ and $K_2$, $a={-1\over{2\tau_{ef}}}$ and $b={\sqrt{4J_{rad}-1}\over{2\tau_{ef}}}$.  Also, in the long time limit, it is possible to show that Eqs. (\ref{x2c}) and (\ref{v2c}) become
\be 
\langle x^2(t)\rangle=2D\left(t+{J_{rad}-3 \over 2}\tau_r\right) , \qquad  \langle v^2\rangle_{eq}={k_BT\over m}.
\label{ltlc} \ee
It is now clear that in this case, $J_{rad}$ can take the threshold value ${\mathcal J}=3$, and therefore the MSD reduces to the Markovian diffusion process $\langle x(t)^2\rangle=2Dt$. 
Again, in the absence of the radiation reaction force,  the MSD and MSV are, respectively, the same as Eqs (\ref{x2c}) and (\ref{v2c}), just replacing $\tau_{ef}$ by $\tau$ and $J_{rad}$ by $J_{bm}$. The same occurs with the MSD in long time. For the threshold value $J_{rad}={\mathcal J}=3$ the evolution of MSD given by Eq. (\ref{x2c}) must be the same and that given in the absence of the radiation reaction force, this is because in this case $J_{rad}=\tau_{ef}/\tau_r=3=\tau/\tau_r=J_{bm}$, which implies that $\tau_{ef}=\tau$, and thus $\tau_0=0$.  As time increases, the MSD follows the same Markovian behavior $2Dt$. The evolution of the MSD is shown in Fig. \ref{Fig.3} and compared with the numerical simulation of Eq. (\ref{sale}). The plot confirms our theoretical predictions.

\begin{figure}
    \centering
    \includegraphics[width=0.55\linewidth]{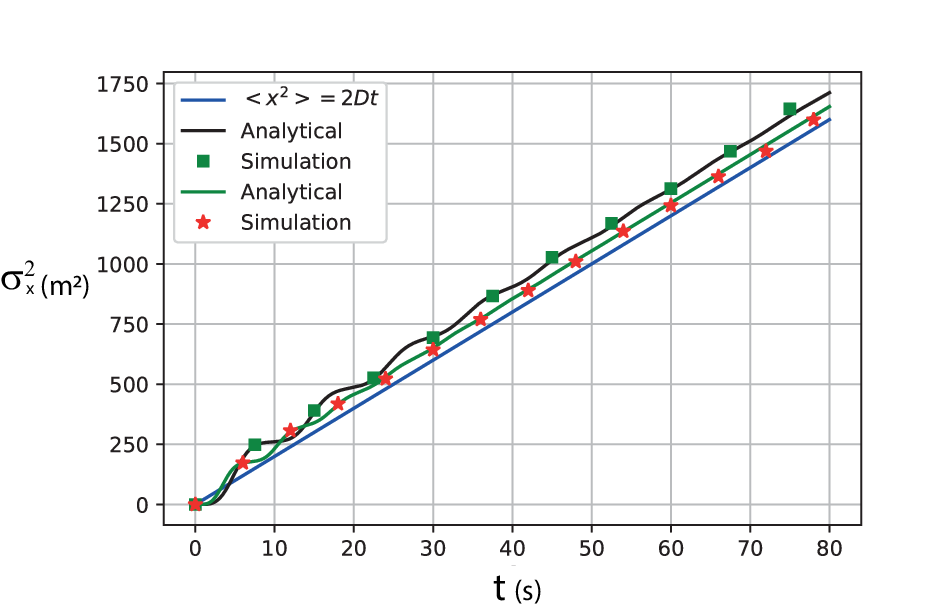}
    \caption{MSD in the case of complex roots. Blue line is the  is the Markovian MSD. Black line corresponds to analytical result (\ref{x2c})  without the radiation reaction force
    ($J_{rad}$ is replaced by $J_{bm}$), and 
    green squares the corresponding numerical simulation of Eq. (\ref{sale}). Green line is the analytical result of Eq. (\ref{x2c}) with radiation reaction force, and red stars the numerical simulation of Eq. (\ref{sale}). The used parameters are:  $J_{bm}=60, ~~ \gamma_0=5, ~~ \tau=12$, ~~   $J_{rad}=30, ~~ \tau_{ef}=6$.}
    \label{Fig.4}
\end{figure}

According to the results shown in Fig. \ref{Fig.3}, one should expect that for values of the parameter $J_{rad}$ greater than the threshold value ${\mathcal J}=3$, ($J_{rad}>3$), the effects of the radiation reaction force on the electronic diffusion process begin to play an important role, i.e., the case for which $\Delta J=J_{bm}-J_{rad}={\tau_0\over \tau_r} >0$. This fact can be corroborated upon the comparison of the analytical results given for the MSD, with and without the presence of the radiation reaction force. The comparison is shown in Fig. \ref{Fig.4}, where the time behavior of both MSDs is plotted and compared with the numerical simulation of Eq. (\ref{sale}); indicates excellent agreement. As time increases, the oscillatory behavior of both MSDs becomes attenuated, taking the shape of a straight line as that given in Eq. (\ref{ltlc}), with $J_{rad}=30$ and $J_{bm}=60$.


\begin{figure}
    \centering
    \includegraphics[width=0.55\linewidth]{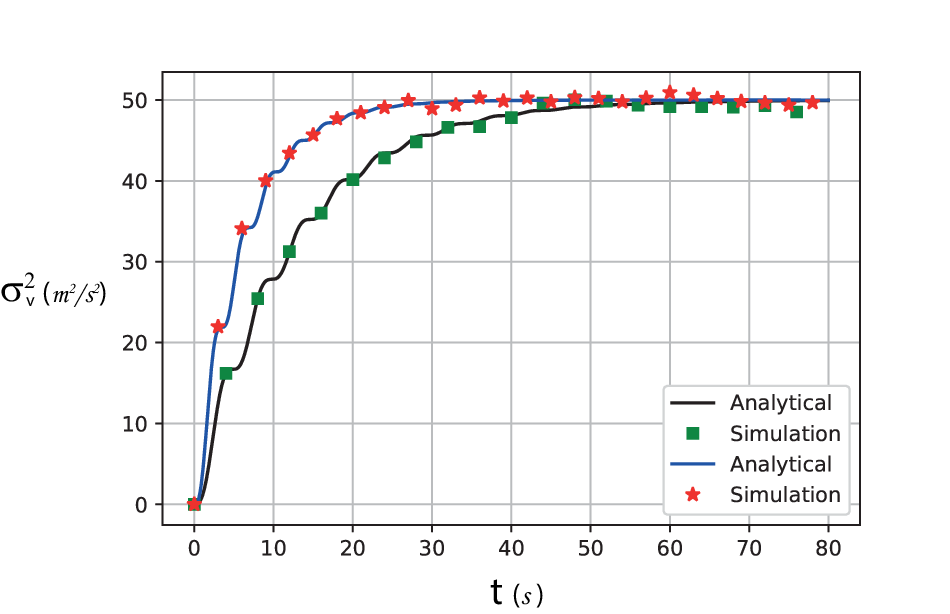}
    \caption{MSV in the case of complex roots. Blue line is the analytical result (\ref{v2c}) without radiation reaction force ($J_{rad}$ is replaced by $J_{bm}$), and red stars the corresponding numerical simulation results. Black line is the analytical result of  Eq. (\ref{v2c}) with the radiation reaction force, and green squares the simulation results. Again $J_{bm}=60$, ~~$\gamma_0=5$, ~~$\tau=12$, ~~$J_{rad}=30$, ~~$\tau_{ef}=6$.}
    \label{Fig.5}
\end{figure}

Lastly, the expression for the MSV has explicitly been obtained for the critical, real, and complex roots. It is shown in each case that as time increases, it becomes the equilibrium result  $\langle v^2\rangle_{eq} =k_BT/m$ as expected. In particular, in Fig. \ref{Fig.5}, the time evolution of the MSV given by Eq. (\ref{v2c}) (black line) is plotted and compared with the numerical simulation of Eq. (\ref{sale}) (green squares). 
In the same Fig. \ref{Fig.5}, the MSV in the absence of the radiation reaction force ($J_{rad}=J_{bm}$) is plotted (blue line), and compared with the numerical simulation of Eq. (\ref{sale}) (red stars). The simulation results given in the plot of Fig. \ref{Fig.5}, have been obtained from the SALE (\ref{sale}) in terms 
of the velocity variable, that is, $m\tau_{ef} \ddot v + m\dot v+\gamma_0 v=\sqrt{\lambda}\, \xi(t)$. The numerical calculations performed to obtain the MSDs rely upon the SALE (\ref{sale}). All the simulation results are given using  Heun's algorithm \cite{Chapra2011}.

\section{Concluding Remarks}\label{sec4}

From a classical statistical mechanics point of view, we have proposed an alternative strategy to characterize electronic plasma diffusion, taking into account the radiation reaction force. Our proposal consists in assuming that the memory time appearing in Kubo's second FDT
(\ref{fdt1}) is an effective memory time defined as $\tau_{ef}=\tau-\tau_0>0$, which accounts for the thermal interaction between the BP and its surroundings. This fact allows us to describe electron Brownian motion by means of a SALE given by Eq. (\ref{sale}). Owing to the positive character of the effective friction force, $m\tau_{ef}\,{\dddot x}>0$, the solution of the SALE cannot violate causality. Our proposal establishes a threshold value of $J_{rad}={\mathcal J}_{rad}=3$, above which the effects of the radiation reaction force on the electronic diffusion process are important. Below the threshold value, such effects have no influence. The consistency of our analytical results can be seen when they are compared with the numerical simulation of SALE (\ref{sale}), showing excellent agreement.     






\vskip0.2cm
\noindent{\bf Acknowledgments}

J. F. G. C. thanks to SECIHTI for a grant in a Postdoc position at UAM campus Iztapalapa Metropolitan Autonomus University. O. C. V. thanks SECIHTI for the grant. G. A. P. and N. S. S. thank COFAA, EDI IPN and SNII-SECIHTI. J. I. J. A.
also thanks to UAM campus Iztapalapa and SNII-SECIHTI.


\end{document}